\documentclass{Styles/sig-alternate-05-2015}

\usepackage{enumerate}
\usepackage{amsmath}

\makeatletter
\def\@copyrightspace{\relax}
\makeatother

\begin{document}

\title{Analyzing Similarity in Mathematical Content \\
      To Enhance the Detection of Academic Plagiarism \\}

\numberofauthors{1}
\author{
\alignauthor
Maurice-Roman Isele\titlenote{maurice-roman.isele@uni-konstanz.de}\\
       \affaddr{University of Konstanz}\\
       \affaddr{Germany}\\
}

\maketitle
\begin{abstract}
Despite the effort put into the detection of academic plagiarism, it continues to be a ubiquitous problem spanning all disciplines. Various tools have been developed to assist human inspectors by automatically identifying suspicious documents. However, to our knowledge currently none of these tools use mathematical content for their analysis. This is problematic, because mathematical content potentially represents a significant amount of the scientific contribution in academic documents. Hence, ignoring mathematical content limits the detection of plagiarism considerably, especially in disciplines with frequent use of mathematics.

This paper aims to help close this gap by providing an overview of existing approaches in mathematical information retrieval and an analysis of their applicability for different possible cases of mathematical plagiarism. I find that whereas syntax-based approaches perform particularly well in detecting undisguised plagiarism, structure-based and hybrid approaches promise to also detect forms of disguised mathematical plagiarism, such as plagiarism with renamed identifiers. However, more research in this area is needed to enable the detection of more complex mathematical plagiarism: the scope of current approaches is restricted to the formula-level, an extension to the section-level is needed. Additionally, the general detection of equivalence transformations is currently not feasible. Despite these remaining problems, I conclude that the presented approaches could already be used for a basic automated detection system targeting mathematical plagiarism and therefore enhance current plagiarism detection systems.
\end{abstract}

\keywords{math similarity; similarity factors; plagiarism detection; math content; math formulae;  math retrieval; math search }

\section{Introduction}
\subsection{Motivation}
The field of mathematical information retrieval has seen significant research progress in recent years, however, compared to traditional text-based information retrieval, a lot of research still has to be done. In particular, assessing the similarity of mathematical content, which is one of the fundamental problems in mathematical information retrieval, still is subject to current research. Many challenges emerge, because of the abstract nature of mathematical expressions and their ambiguous representation: In addition to syntax and semantics, notational ambiguities must also be considered in the similarity analysis. Furthermore, mathematical content does not only occur as single isolated formulae or expressions, but it can stretch over several paragraphs using many mathematical or textual fragments (for example proofs). It is desirable to not only analyze those isolated fragments, but to define a similarity measure that includes all those fragments in the analysis for a combined similarity. This, however, is a very challenging and complex task together with the difficulties mentioned above.

To our knowledge, current techniques for plagiarism detection do not analyze the similarity of mathematical content. Popular text-based approaches include string matching, where documents are compared for text overlaps, or measuring the similarity of documents via bag of words and a similarity measure such as the cosine similarity. This however does not cover translated or paraphrased text, therefore this technique is not sufficient. More sophisticated approaches have been proposed, such as citation-based plagiarism detection \cite{gipp2014citation}. This technique searches for similar citation patterns and can thus even detect plagiarism in translated or paraphrased text. But since the analysis requires properly cited sources, it cannot cope with documents where citations are incorrect, incomplete, or nonexistent.

However, adding mathematical content into the scope of the analysis would be highly beneficial: For one, the weaknesses of the text matching techniques mentioned above do not directly apply to mathematical content: Math is a universal language and can therefore be analyzed independently from translated text. Thus, this approach also works for translations. Additionally, although there are various ways to disguise mathematical plagiarism, the deeper structure of the mathematical content remains the same. Otherwise, the original and the disguised content could not express the same meaning. Hence, paraphrased mathematical content can still be detected. Also, in contrast to citation-based plagiarism detection, the detection of plagiarized mathematical content does not rely on correct and complete citations, the analysis of the content itself is sufficient. Most importantly, mathematical content potentially represents a significant amount of the scientific contribution, hence ignoring it limits the detection of plagiarism considerably, especially in disciplines with frequent use of mathematics.

For these reasons, including mathematical content into the scope of automated plagiarism detection systems is a task worth the effort, despite the challenges already mentioned.

The idea to analyze mathematics to detect plagiarism has recently been proposed by Meuschke et al.\ in \cite{Meuschke:2017:AMC:3132847.3133144}.
The authors applied basic MIR methods to demonstrate the potential of the approach. Their results are promising, yet leave much room for more in depth research. In this paper, I will lay a small foundation for future research in this area. For this purpose, I will review existing methods from mathematical information retrieval and analyze their strengths and weaknesses in regard to plagiarism detection in Section Two. Then, I will discuss their performance for various possible cases of mathematical plagiarism and highlight open research problems in Section Three. Finally, I will summarize my findings in the Conclusion and give an overview of the tasks that need to be considered in further research in the Outlook.

\subsection{Research Questions}

\noindent In this paper, I investigate the following research question:\\

{
\centering{\textit{
''How can similarity assessments of mathematical content \\
aid in the detection of academic plagiarism?'' \\}}
}
~\\

\noindent As already outlined above, we tackle this question by dividing it into two research tasks, which represent the further structure of this paper:

\begin{enumerate}[i)]
\item Determining the similarity of mathematical content
	\begin{itemize}
	\renewcommand\labelitemi{-}
	\item Review of existing approaches in MIR
	\item Pros/Cons in regard to our use case
	\end{itemize}
\item Analyzing mathematical plagiarism
	\begin{itemize}
	\renewcommand\labelitemi{-}
	\item Formalizing possible variations
	\item Analyzing the applicability of the presented approaches
	\item Highlighting open research problems
	\end{itemize}
\end{enumerate}

\noindent where i) and ii) will be examined in Section Two and Three respectively.

\section{Determining Math Similarity}

In this Section, I examine research task \textit{(i)}. For this purpose, I review existing and exemplary approaches from mathematical information retrieval that try to determine the similarity of mathematical content. For every approach, I elaborate advantages and disadvantages in regard to plagiarism detection.

Researchers agree that traditional MIR approaches can be categorized into syntax-based approaches and structure-based approaches \cite{yokoi2009approach,kohlhase2006search,zhang2014approach}. Syntax-based approaches focus on the similarity of the tokens representing the mathematical expression, whereas structure-based approaches focus on the structural similarity of mathematical expressions. Additionally, I found that there are approaches which combine syntax-based and structure-based features, which I will refer to as ''hybrid approaches''. Finally, there are semantic approaches, which try to capture the semantic context of mathematical expressions. The body of this Section follows this categorization into syntactical, structural, hybrid, and semantic approaches.

\subsection{Syntax-based approaches}

Syntax- or text-based approaches try to make use of the already sophisticated knowledge in traditional text retrieval. The core idea is to measure the similarity of the strings representing the mathematical content. For this purpose, syntax-based approaches convert mathematical expressions into a format that can be used in text retrieval systems. By building upon existing and mature techniques instead of building a mathematical search engine from scratch syntax-based approaches save effort. Usually, formats such as linear strings or bags of words are used. Miller and Youssef demonstrate a possible conversion in \cite{miller2003technical}. Using what they call \textit{''Textualization''} and \textit{''Flattening''} a result like the following is obtained: A mathematical expression such as ''$x^{t+2} = 1$'' is converted into a text-based format like

\begin{center}
''\textit{x BeginExponent t minus 2 EndExponent Equal 1}''
\end{center}

\noindent This format is applicable for text retrieval systems, in contrast to the original formula. Various matching techniques can be applied using this format. However, most of these syntax-based approaches do not only translate the mathematical expression into an adequate format and perform a simple matching. Syntax-based approaches usually also add normalization steps to improve performance. Miller and Youssef also perform an ordering of the characters. On every operator that is commutative, they apply an alphabetic ordering, this way the mathematical expression $1 = x^{2+t}$ is transformed into the same format like the one displayed above, which is desirable for both their use case and our use case to detect mathematical plagiarism.

The \textit{Mathdex search engine} \cite{miner2007approach} is another example of a syntax-based approach. This approach is similar to the one presented by Miller and Youssef, however the Mathdex search engine adds more sophisticated normalization steps which are very useful for plagiarism detection:
\begin{enumerate}
\item \textit{Notational normalization:}
Redundant white-spaces are being removed (e.g. ''$x ~ + ~ 2 ~ = ~ 1$'' is reduced to ''$x + 2 = 1$''). Also, similar characters with a different Unicode are being normalized.

\item \textit{Mathematical normalization:}
Synonymous representations for the same mathematical construct are being normalized. For example, 
''n choose k'' can be written as (among others): $n \choose k$ , $C(n,k)$ , $~_nC_k$ , \dots
\item \textit{Variable names:} The search engine does not only search for the exact query, but also tries to replace variable names with other common identifiers (e.g. i,j,k for sums). For example, if the query contains a sum ''$\sum\nolimits_{a=1}^n{a^2}$'', then the engine will also consider ''$\sum\nolimits_{i=1}^n{i^2}$''.
\end{enumerate}

Changing the notation, using synonymous representations, or changing variable names are popular ways to hide plagiarism. Therefore, these normalizations are very interesting for our use case. The normalization steps 1. and 2. are effective. The normalization step 3. is a valuable starting point for basic, frequently occurring variables, yet leaves room for future extensions because it is limited to specific mathematic constructs (e.g. sums) and common identifiers. The change of variable names, however, is possible at arbitrary positions and with arbitrary identifiers. Therefore, it is desirable that the engine is not restricted to a number of mathematical constructs or common identifiers.\\

\begin{figure}
\includegraphics[scale=0.3]{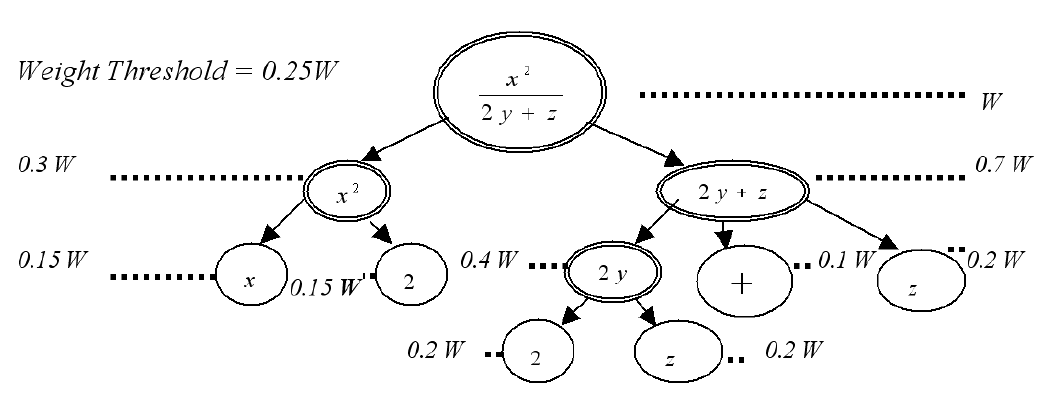}
\caption{Weighted query tree \newline
Illustration taken from \cite{miner2007approach}}
\end{figure}

\noindent The \textit{Mathdex search engine} furthermore extends the literal matching by using weighted n-gram trees (see figure 1). A mathematical expression is transformed into n-grams, to which a weight is assigned depending on its complexity. Then, the engine does not only search for the full query, but also for all n-grams which satisfy a certain weight threshold. The weight of the n-gram determines how much this n-gram contributes to the overall similarity score. This technique allows for a more intuitive matching of sub-terms: a more complex term should be more important for the engine, than a simple term. For our use case this might be important if, for example, plagiarized material is hidden as a sub-term inside a bigger formula.

Further syntax-based approaches, such as \textit{ActiveMath} \cite{libbrecht2006methods} are similar to the two previously presented. They all share the core idea to use existing text-retrieval engines. For their similarity measure they focus on the string representing the mathematical content, adding normalization steps or techniques like n-grams for a more sophisticated matching.\\

\noindent The main advantage of syntax-based approaches is that they are straightforward and realizable with little effort. The matching technique is intuitive: the more of a query is covered in a formula, the more similar they are. Syntax-based approaches also seem to be ideal for the detection of plagiarized mathematical content, because they are inherently good at detecting direct matches (i.e. copy \& paste plagiarism). Due to the sophisticated normalizations, even slightly altered mathematical content is detectable. However, syntax-based approaches do not consider structural information for retrieving similar mathematical content and only a fraction of the semantics is captured. That's why detecting more complex plagiarism, such as paraphrased formulae or the change of variable names, is problematic when using this technique. Also, there is a risk of many false positives when the engine is querying small fractions of mathematical content: Expressions like ''(Let) $x=1$'' may appear in many documents, and in many cases this is not relevant for our use case. Additionally, common formulae such as ''$E = mc^2$'' lead to the same problems. They appear in many academic documents in physics, but it is highly unlikely that an author using this formula is plagiarizing, since this formula is common knowledge in physics.

\subsection{Structure-based approaches}

Structure-based approaches are considerably different from syntax-based approaches. They concentrate on capturing the structural information of the mathematical expressions. For this purpose, structure-based approaches use markup languages, such as \textit{XML}. Generally, they transform mathematical expressions into trees and calculate the similarity score on those trees, rather than on the string representing the mathematical expression like the syntax-based approaches do.

\begin{figure}
\includegraphics[scale=0.4]{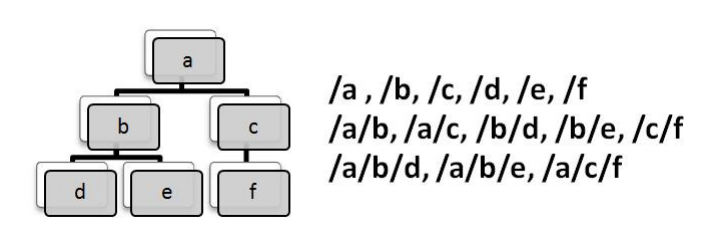}
\caption{Example of a tree and its Subpath Set.
Illustration taken from \cite{yokoi2009approach}}
\end{figure}

\noindent The approach by Keisuke Yokoi and Akiko Aizawa presented in \cite{yokoi2009approach} is an example. For their similarity measure they use \textit{''Subpath Sets''}, which was initially proposed by Ichikawa et al. \cite{ichikawa2005new}. A Subpath Set is defined as \textit{''the path from the root to the leaves and all the sub-paths of that''} (see figure 2), and can be used to assess the similarity of trees. \\

\noindent They calculate their similarity measure as follows ($t_i$ refers to a tree and $S(t_i)$ to the Subpath Set of that tree):

$$ sim(t_1,t_2) = \frac{||S(t_1) \cap S(t_2)||}{||S(t_1) \cup S(t_2)||}  $$
~

\noindent Hence, the degree of overlap of the Subpath Sets determines the similarity of the corresponding trees.\\

\begin{figure}
\includegraphics[scale=0.9]{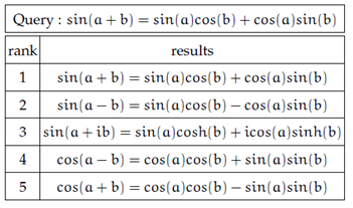}
\caption{Exemplary results of a query search \newline
Table taken from \cite{yokoi2009approach}}
\end{figure}

 \noindent Yokoi and Aizawa find that their method \textit{''is capable of evaluating the structural similarities of the trees rather than the notational similarities of the tokens''}. Figure 3 demonstrates an exemplary query and the corresponding search results. All of the results seem to be very similar, the overall structure of the expressions is the same. The top result is a direct match, as hoped-for. However, despite being structurally similar, semantically the results are significantly different from each other. It's questionable whether we would be interested in each of these results when searching for mathematical plagiarism. Therefore, this approach might lead to false positives in our use case.\\

\noindent \textit{MathWebSearch}, presented by Michael Kohlhase and Ioan A. Sucan in \cite{kohlhase2006search}, is another example of a structure-based approach. The search engine
works with \textit{MathML} and \textit{OpenMath}. These are markup languages used for the representation of mathematical formulae. The engine uses normalization steps similar to the ones presented in the previous Subsection and uses substitution trees to process the \textit{MathML}- and \textit{OpenMath}-objects.

\textit{MathML} can be subdivided into \textit{Presentation MathML}, which focuses on the layout and display of formulae, and \textit{Content MathML}, which focuses on the semantics of formulae. The additional semantics provided by \textit{Content MathML} allowed Kohlhase and Sucan to implement $\alpha-equivalence$. This means that \textit{MathWebSearch} considers two formulae represented in \textit{Content MathML} to be equivalent even if bound variables have been renamed. This is a significant advantage over the syntax-based approaches, which are unable to support $\alpha-equivalence$. This is also very interesting for our use case, because the renaming of variables is a common way to disguise plagiarism. However, \textit{MathWebSearch} supports $\alpha-equivalence$ only for formulae represented in \textit{Content MathML}. This can be a very limiting factor, since, as Kohlhase and Sucan state: \textit{''content representations are often hidden in repositories, only their presentations are available on the web''}. Another disadvantage of \textit{MathWebSearch} is that it does not support searching for simple text, therefore it cannot process constructs like proofs which contain textual fragments. Also, popular formats such as \LaTeX ~ are not supported. \\

\noindent To summarize, structure-based approaches are able to capture structural similarities of mathematical expressions, in contrast to syntax-based approaches which loose the structural information during their conversion process. \textit{MathWebSearch} shows how structure-based approaches are capable of adding more semantics into the analysis.

This is very promising for more complex mathematical plagiarism, for example where variable names are changed. Syntax-based approaches are unable to detect arbitrary renaming of variables. Hence, structure-based approaches have a decisive strength in this regard.

Also, they usually perform well in highly ranking direct matches according to the evaluations in \cite{yokoi2009approach}. However, there is a danger of false positives, because the top results tend to include mathematical expressions that are structurally similar to the query, but semantically very different and therefore most likely not interesting for a plagiarism detection system.

\subsection{Hybrid approaches}
\begin{figure}
\includegraphics[scale=0.325]{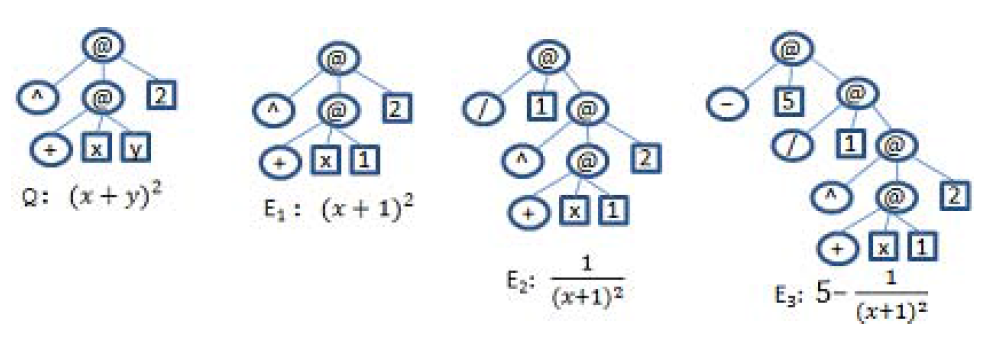}
\caption{Match-Depth example\newline
For a Query $Q$ and Equations $E_1$, $E_2$, $E_3$ \newline
Illustration taken from \cite{zhang2014approach}}
\end{figure}

Hybrid approaches try to combine the ideas and strengths of both syntax-based, and structure-based approaches. One example of such an approach is the math-similarity metric published by Qun Zhang and Abdou Youssef in \cite{zhang2014approach}. They proposed a similarity measure using five factors:
\begin{itemize}
\item \textit{Taxonomic Distance of Functions}, which increases similarity of two terms belonging to the same category according to a content dictionary (e.g. ''+'' and ''-'' may be more similar than ''+'' and ''cosinus'' depending on the content dictionary),
\item \textit{Data Type Hierarchy Level}, which treats elements as more similar when they are hierarchically closer (e.g. comparing two variables or a variable with a function),
\item \textit{Match-Depth}, which decreases similarity if a term is nested deeper inside a formula (see figure 4),
\item \textit{Query Coverage}, which takes into account how much of the formula is covered,
\item and finally \textit{Formula vs.\ Expression}, which considers more relevance to full formulae.
\end{itemize}

As pointed out in an evaluation \cite{schubotz2014evaluation}, the first four factors are an effective measure to assess mathematical similarity (in the context of math search). The last factor is not as important. Presumably, this is also true for plagiarism detection, although it would be interesting to experiment by applying different weightings to each factor.

Similar to the structure-based approaches, Zhang and Youssef use trees to represent mathematical expressions. They focus on mathematical expressions encoded in \textit{(Strict) Content MathML}. Using their five factors they calculate the similarity recursively via the height of the trees. The factors and the recursive calculation contain multiple parameters which can be altered to change the behavior of the similarity metric. For example, it is possible to use a linear, quadratic, or even exponential decay factor for the \textit{Match-Depth}. This way, one can manipulate the approach to fit a specific task. However,
the choice of the parameters is not necessarily intuitive, and one needs to experiment to find the optimal choice. As Zhang and Youssef state, the parameter values need yet to be optimized. \\

A significant advantage of this approach is that it combines the strengths of syntax-based and structure-based approaches. Structural, syntactical, and semantic information is captured by this approach.  According to the evaluation of Zhang and Youssef, their approach performs well compared to the \textit{DLMF (Digital Library of Mathematical Functions)} \cite{NIST:DLMF}. Also, the parameters for each factors allow for an amplification or weakening of single factors, which makes it possible to tailor this similarity measure to a specific use case. However, those parameters are also problematic: To obtain satisfying results, these parameters must be optimized. Additionally, the influence of many factors makes the analysis in Section Three more difficult.

Another advantage of this similarity measure is that it results in a value between 0 and 1. Hence it allows for a relevance ranking which can be easily modified into a ''degree of suspiciousness'' in regard to plagiarism detection, which is essential for an automated plagiarism detection system.

\subsection{Semantic approaches}

Semantic approaches try to determine the semantic context of mathematical expressions. For example, mathematical expressions containing an identifier ''E'' may refer to \textit{Energy} in physics or \textit{Expectancy value} in statistics. The goal of semantic approaches is to remove these ambiguities. One example for such an approach was recently proposed by Schubotz et. al in \cite{schubotz2016semantification}.\\

In their approach, they determine the semantic context of identifiers. They do so by searching for definitions in the text near mathematical expressions. For example, if there is a formula like ''$E = mc^2$'' they search the surrounding text for definitions of the form ''\textit{... where E is energy}'' or ''\textit{c is the speed of light}''. Then, they use their knowledge to disambiguate identifiers by adding a corresponding subscript. In this example, their approach would transform ''$E$'' to ''$E\_{energy}$''.\\

The disambiguation of identifiers allows for the reduction of false positives in both math search and plagiarism detection. If someone is interested in a formula containing ''E'' in the context of energy, then results from statistics that refer to the expectancy value are irrelevant. Likewise for plagiarism detection: It is highly unlikely that a formula in a paper about statistics referring to the expectancy value was copied from a physics paper referring to energy.\\

Hence, this approach might be used to augment the previously presented approaches. This approach could help reducing false positives and the total number of documents a plagiarism detection system has to consider. This way, less resources would be needed by the plagiarism detection system.

\section{Mathematical Plagiarism}

In this section, I illustrate possible variations of mathematical plagiarism and elaborate on emerging challenges for plagiarism detection systems. Then, I analyze the applicability of the approaches presented in Section Two and highlight open research problems.

The variations of mathematical plagiarism presented below are ordered by the complexity of their disguise, i.e. from undisguised plagiarism to increasingly sophisticated plagiarism.

\begin{figure}
\centering
\includegraphics[scale=0.55]{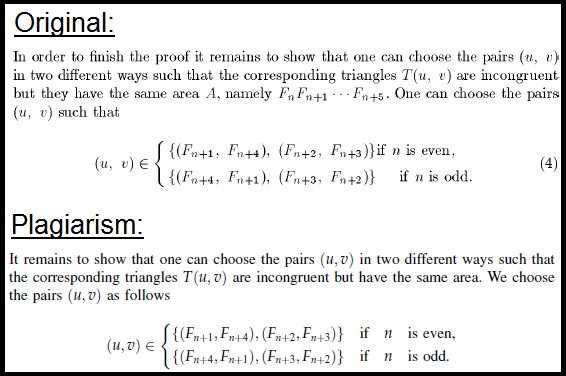}
\caption{Plagiarism using simple copy \& paste. \newline
Original: ''\textit{Some remarks on Heron triangles}'' \cite{kramer2000some} }
\end{figure}

\subsection{Simple Copy \& Paste}

\subsubsection{Definition and Challenges}

The most basic form of mathematical plagiarism is 
''Copy \& Paste''-plagiarism, i.e. undisguised plagiarism, where the plagiarized content is an exact copy of the original content. An example of such a case can be seen in Figure 5: The excerpts from the original document and the plagiarized document contain the exact same mathematical expression. The plagiarists slightly altered the surrounding text, but decided not to paraphrase the formula. In such a case, it would be sufficient for a plagiarism detection system to only find exact matches. A plagiarism detection system able to detect undisguised plagiarism would force plagiarists to not only paraphrase text, but to paraphrase mathematical expressions. This way, more effort would be needed to successfully hide mathematical plagiarism.\\

However, trying to detect this kind of plagiarism leads to a major challenge: In many fields, various formulae are considered common knowledge and therefore (legally) ''copied'' (for example $E = mc^2$). Hence, a naive approach could possibly return many matching, but non-plagiarized formulae (i.e. false positives). This would hurt usability decisively, because human inspectors would then have to examine many irrelevant matches. Differentiating between a ''Copy \& Paste''-plagiarism and the usage of a common knowledge formula is not trivial, but could be tackled by identifying and analyzing citations (if present) or using a ''whitelist''. Another possible approach, as presented in \cite{gipp2014citation}, is to assign less weight to formulae in the introduction and related-work sections, since such content likely represents existing knowledge.\\

\subsubsection{Applicability of Presented Approaches}

Finding exact matches is sufficient for the detection of undisguised plagiarism. Therefore, all of the presented approaches should be capable of detecting this instance of mathematical plagiarism.

However, syntax-based approaches promise the best performance, because they are inherently good at finding direct matches. Structure-based approaches should perform worse than syntax-based approaches, because analyzing the structure of the mathematical expressions is not necessary for the detection of undisguised plagiarism. Structure-based approaches tend to produce more false positives that are similar in structure, but irrelevant for plagiarism detection. The performance of the presented hybrid approach strongly depends on the choice of parameters, but it should be similar to the performance of the syntax-based approach. Including the semantic approach into the analysis can help reducing false positives for the other approaches, as already elaborated in Section Two.

\subsection{Notational changes}

\subsubsection{Definition and Challenges}

An easy and common way to disguise mathematical plagiarism is to make simple notational changes. This way, the plagiarism is less obvious, both to human inspectors and plagiarism detection systems. I distinguish between two cases of notational changes, which I define as follows:\\

\begin{enumerate}
\item \textbf{Using different characters:}\\
The idea here is to replace characters from the original content with characters that appear similar to a human reader, but are perceived as different by the computer. For example, there are multiple unicode characters for the minus sign: 
among others, there is ''\textit{U+2212}'' and ''\textit{U+002D}'', i.e. two different unicodes for the exact same character. However, the replaced characters do not necessarily have to look the same as the original characters. It is sufficient if they are perceived as equal by a human reader. For example, one could use ''$\cdot$'' or ''$*$'' for the multiplication sign (or even omit it).

\item \textbf{Using different representations:}\\
The same idea is also applicable for mathematical constructs. Often, there are many synonymous representations for the same concept. An example is the binomial coefficient, which has many synonymous representations:
$$\frac{n!}{k!(n-k)!} = \binom{n}{k} = C(n,k) = {}_{n}C_k = C_k^n = \dots$$

\noindent A more complex example for notational changes is using a different representation for fractions:

$$\frac{a + b + c}{n} = \frac{a}{n} + \frac{b}{n} + \frac{c}{n}$$

\end{enumerate}

\noindent For the detection of mathematical plagiarism with notational changes, it is no longer sufficient to only find exact matches. In this case, normalizations are needed like the ones used by the \textit{Mathdex search engine} presented in Section 2.1. However, normalizing all possible notational changes is a difficult task, because plagiarists tend to be creative in finding new exploits. Additionally, although the normalization of commutative terms is already used in current math search engines, it is a more complex task to normalize distributive terms (as in the example of multiple notations for fractions). Still, many notational changes can be tackled by adding appropriate normalization steps to the plagiarism detection system.

\subsubsection{Applicability of Presented Approaches}

If we assume sufficient normalization, then all the notational changes get nullified and the detection of this instance of plagiarism is again reduced to finding direct matches. In this case, we can expect results similar to the ones elaborated for Copy \& Paste - plagiarism.

However, as already addressed in the Section \textit{Definition and Challenges}, it is not trivial to implement a normalization pipeline that is sophisticated enough to cover all possible notational changes. Therefore, it might be more realistic to assume that the normalization is not complete. This would be problematic, especially for the syntax-based approaches, which would have difficulties matching notational changes remaining after the normalization step. Structure-based- and hybrid-approaches, on the other hand, should perform better, because they don't rely as much on literal matching.

\newpage

\subsection{Renaming identifiers}

\subsubsection{Definition and Challenges}

\begin{figure}
\begin{center}
\includegraphics[scale=0.4]{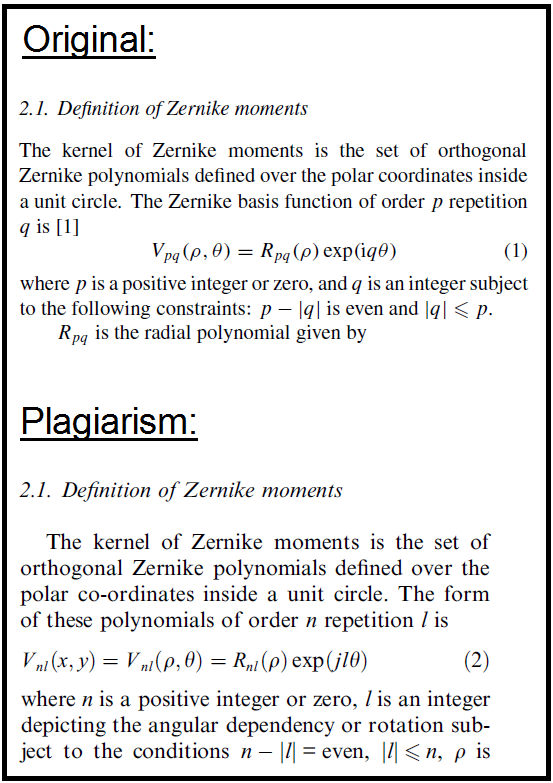}
\end{center}
\caption{Plagiarism with renamed identifiers. \newline
Original: ''\textit{Invariance analysis of improved Zernike
moments}'' \cite{bin2002invariance} 
}
\end{figure}

In addition to simple notational changes, plagiarists often try to disguise their plagiarism by renaming identifiers. Figure 6 illustrates an example of such a case. Here, the plagiarists renamed $ p,q,i $ to $ n,l,j $.

Unlike simple notational changes, the renaming of identifiers cannot be reasonably tackled by normalization techniques. The presented normalizations by the \textit{Mathdex Search engine} contain some normalizations for identifier names, however, these are restricted to specific constructs (e.g. sums) and common identifiers (e.g. $ i,j,k $). Normalizing general identifier renaming is not possible, because there are too many possibilities for different identifier names. Even if concepts like wild-cards are introduced, it is not possible to normalize all legal renamings of expressions such as '$a + b = a^2$', because wild-cards cannot capture that both 'a's must be renamed together, i.e. '$x + y = z^2$' is no legal renaming of the expression above.

\subsubsection{Applicability of Presented Approaches}

Detecting plagiarism with renamed identifiers is problematic and generally not possible for syntax-based approaches, because of the aforementioned challenges.

This is however, where the structure-based approaches show their strengths: by focusing on the structure of the mathematical expressions rather than on the tokens, they find results that are structurally similar, independent of any renamed identifiers. Approaches like \textit{MathWebSearch} even implement $\alpha$-equivalence, which is exactly what we need for this scenario.

Since the hybrid approach has an additional structural component, it should perform better than the syntax-based approach. However, the performance strongly depends on the parameters and is difficult to predict.

\subsection{Splitting content}

\subsubsection{Definition and Challenges}

Apart from changing single fragments of the original mathematical content, it is also possible to change how the content is represented as a whole. For instance, plagiarists could split the content into multiple parts, possibly adding additional text. Consider the following simple, constructed example (this is the induction step for \textit{Bernoulli's inequality}):
\begin{itemize}
\item \noindent \underline{\textbf{Original:}}\\
\begin{align*}
(1+x)^{t+1} &= (1+x)^t \cdot (1+x)\\
            &\geq (1+tx)(1+x)\\
            &= 1 + x + tx + tx^2\\
            &\geq 1 + (t+1)x
\end{align*}
\item \noindent \underline{\textbf{Plagiarism:}} \\
\begin{align*}
(1+x)^{t+1} &= (1+x)^t \cdot (1+x)\\
\end{align*}
Because of $x > -1$ we can apply the induction hypothesis and can conclude that 
\begin{align*}
(1+x)^{t+1} &\geq (1+tx)(1+x)\\
            &= 1 + x + tx + tx^2\\
            &= 1 + (t+1)x + tx^2
\end{align*}
Finally, since $tx^2 \geq 0$, we get
\begin{align*}
(1+x)^{t+1} &\geq 1 + (t+1)x\\
\end{align*}
\end{itemize}

\noindent By splitting the formula into multiple parts, it is more difficult for plagiarism detection systems to detect copied content. This is because an extension from the formula- to the section-level is needed for this instance of plagiarism, i.e. it would be necessary to recognize that the formulae in the plagiarism-example belong together. 

\subsubsection{Applicability of Presented Approaches}

Since all of the presented approaches work on the formula-level, the similarity score of all approaches is decreased if mathematical content is split. Therefore, detecting this instance of mathematical plagiarism is problematic for all of the presented approaches. As already mentioned, the approaches would have to be extended to not only work with isolated formulae, but with entire sections.

\subsection{Introducing/Reducing intermediate steps}

\subsubsection{Definition and Challenges}

Instead of splitting mathematical content, it is also possible to introduce or reduce intermediate steps. This way, the ratio between the suspicious material and the total material is reduced. Therefore, the document as a whole might be classified as less suspicious or even not suspicious as a result.

However, plagiarized material which has not been reduced can still be detected, and introducing/reducing intermediate steps has limits: Cutting too many intermediate steps leads to a loss of contiguity and substance, introducing too many intermediate steps lowers the quality of the content and might hinder publishing of the paper.\\

For this instance of mathematical plagiarism the following question arises: when should a document be marked as ''suspicious'' and passed on to the human inspector? I.e., how should the threshold for the degree of suspiciousness be defined? This is an important question. If the threshold is too high, then documents which contain plagiarism might be marked as ''not suspicious''. If, however, the threshold is too low, then the human inspectors will have to review (too) many documents. Experimenting and experience is necessary to find the optimal choice for the threshold.

\subsubsection{Applicability of Presented Approaches}

The described challenges are independent from the presented approaches.\\

\subsection{Substituting terms}

\begin{figure}
\begin{center}
\includegraphics[scale=0.4]{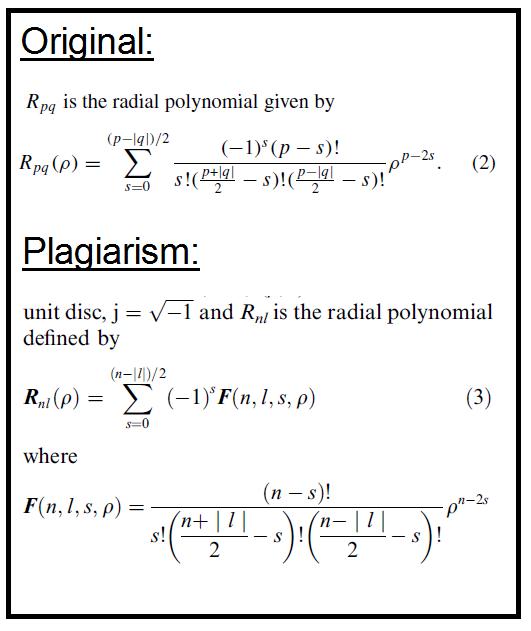}
\end{center}
\caption{Plagiarism with substituted terms. \newline
Original: ''\textit{Invariance analysis of improved Zernike
moments}'' \cite{bin2002invariance} 
}
\end{figure}

\subsubsection{Definition and Challenges}

A more complex form of mathematical plagiarism is the substitution of terms. Figure 7 shows an example. These excerpts are the continuation of the excerpts from Figure 6, that is why the identifier names are renamed. Here, the plagiarists replaced a larger term with a function call.

This problem is similar to the splitting of mathematical content, that is why the extension from formula- to section-level is a problem here too. However in this case the original formula is directly altered as well (it is not a simple split).

\subsubsection{Applicability of Presented Approaches}

The applicability of the presented approaches is very limited for the same reasons as presented for the splitting of mathematical content. However, semantic approaches might offer an interesting solution here. One could imagine an approach that is similar to the presented semantic approach: searching for identifiers, searching for definitions, and then replacing the identifier with the definition. This way, the example in Figure 7 could be solved theoretically. However, no such approach exists yet, and it would likely be costly to add such an approach to the plagiarism detection system.

\subsection{Equivalence Transformations}

\subsubsection{Definition and Challenges}

One of the most complex task for plagiarism detection systems is to detect mathematical plagiarism where equivalence transformations have been applied. In order to disguise their plagiarism, one can perform multiple simple transformations until the plagiarized content is significantly different from the original content. Apart from simple transformations, one can also use transformation rules such as the rules for the logarithm:

\begin{align*}
E_1 &:= \log_a(\frac{x+2}{y^2})\\
& \\
E_2 &:= \log_a (x+2) - \log_a y^2\\
\end{align*}

$ E_1 $ and $ E_2 $ are semantically equivalent, but both syntactically and structurally different.\\

\noindent Whereas transformation rules like in the example can be ''hard-coded'', the recognition of general equivalence transformations would require a math engine. However, this would come with a high computational effort. Considering that a plagiarism detection system usually works on a very large database with many documents, adding a math engine to the detection pipeline is simply not feasible.

\subsubsection{Applicability of Presented Approaches}

None of the presented approaches tackles the problem of equivalence transformations. This is why this instance of mathematical plagiarism is problematic for all of them. Apart from hard-coded rules, this problem can currently not be solved.\\

\subsection{Summary}

\tiny
\begin{center}
	\hspace*{-0.25cm}
	\begin{tabular}{| p{1.2cm} | p{1.1cm} | p{1.1cm} | p{1.1cm} | p{1.1cm} | p{1.1cm} |}
	\hline
	& \textbf{Challenges} & \textbf{Syntactical} & \textbf{Structural} & \textbf{Hybrid} & \textbf{Semantic} \\ \hline
	\textbf{Copy \& Paste} & False positives & Ideal & Possible false positives & Good & Reduce false positives \\ \hline
	\textbf{Notational changes} & Complete normalization & Good & Possible false positives & Good & Reduce false positives \\ \hline
	\textbf{Renaming identifiers} & Can't normalize & Problematic & Ideal & ? & Reduce false positives \\ \hline
	\textbf{Splitting content} & Extension to section-level & Problematic & Problematic & Problematic & Reduce false positives \\ \hline
	\textbf{Introducing / Reducing steps} & Lower degree of suspiciousness &  &  &  & Reduce false positives \\ \hline
	\textbf{Substitute terms} & Extension to section-level & Problematic & Problematic & Problematic & Ideal (theoretically) \\ \hline
	\textbf{Equivalence transformations} & Computa- tional effort & Not feasible & Not feasible & Not feasible & \\ \hline
	\end{tabular}
\end{center}
~
\normalsize

\section{Conclusions}
~

In this paper, I investigated how similarity assessments of mathematical content can aid in the detection of mathematical plagiarism.

For this purpose, the first research task was to review existing approaches from mathematical information retrieval in order to find possible ways to determine mathematical similarity, and how we could use this knowledge in plagiarism detection. I found that there are four general approaches for determining mathematical similarity: First, \\ \textbf{syntax-based} approaches, which concentrate on the similarity of the tokens representing mathematical expressions. These approaches usually work on top of text-retrieval engines and use various normalization steps in order to cope with synonymous representations of mathematical content and characters. These approaches are realizable with little effort, and offer an intuitive matching technique for plagiarism detection. Second, there are \textbf{structure-based} approaches, which focus on structural similarity and use trees to represent mathematical expressions.
By focusing on the structural similarity, they introduce the danger of additional false positives, however they also offer the possibility to tackle more complex plagiarism. Third, there are \textbf{hybrid} approaches. These approaches try to combine the strengths of syntax-based and structure-based approaches, and could therefore be a compromise between the two. Lastly, there are \textbf{semantic} approaches. These try to identify the semantic context of identifiers (and formulae). This way, they could reduce false positives and augment the other approaches.\\

\noindent In Section Three, I tackled the second research task: I formalized possible variations of mathematical plagiarism and analyzed the applicability of the presented approaches. As I have discussed in Section Two, the different approaches have different strengths and weaknesses, likewise I found that some approaches should perform better in certain cases of mathematical plagiarism and worse in others. The general trend seems to be that syntax-based approaches perform very well for more simple plagiarism (Copy\&Paste, Notational Changes), while structure-based approaches can tackle more complex mathematical plagiarism (Renaming identifiers). The hybrid approaches could offer a good trade-off here. The semantic approaches can reduce false positives by ignoring documents that are unlikely to be relevant for the plagiarism detection system. Additionally, they have the capability of detecting plagiarism with substituted terms.
However, many of the more complex cases of mathematical plagiarism have many difficulties that cannot be solved with current techniques (e.g. the extension from formula- to section level).\\

\noindent I conclude that a simple plagiarism detection system for mathematical content could be implemented already today. Depending on the demands of the people using this system, an appropriate approach can be chosen (a system with multiple approaches might be possible too). Although there are many challenges left, especially for complex plagiarism, a first basic plagiarism detection system for mathematical plagiarism would close the current gap in plagiarism detection. It would make it more difficult for plagiarizers to disguise their plagiarism, and therefore hopefully discourage them from even trying.

\section{Outlook}

It seems promising to embed a similarity analysis of mathematical features as a component of an integrated detection process. Research indicates that not a single, but combined PD approaches are most promising to detect the different forms of plagiarism ranging from copy and paste to strongly disguised idea plagiarism \cite{gipp2014citation}. The integrated detection process should analyze literal text matches, academic citations, images, mathematical content as well as semantic and syntactic features \cite{Meuschke2014, Meuschke2017}. Considering different forms of document similarity increases the effort required for obfuscating plagiarism, hence increases the deterrent effect of PD systems and thus helps to prevent plagiarism.

\section{Acknowledgements}

I would like to thank Norman Meuschke and Moritz Schubotz for their valuable feedback.

\bibliographystyle{unsrt}
\bibliography{references}
\end{document}